# Surface Energy Engineering for Tunable Wettability through Controlled Synthesis of MoS$_2$

Anand P.S. Gaur,[1] Satyaprakash Sahoo,[1,*] Majid Ahmadi,[1] Saroj P Dash,[2] Maxime J-F Guinel[1] and Ram S. Katiyar[1,*]

[1]Department of Physics and Institute for Functional Nanomaterials, University of Puerto Rico, San Juan, PR 00931 USA

[2]Department of Microtechnology and Nanoscience, Chalmers University of Technology, SE-41296, Göteborg, Sweden

**MoS$_2$ is one of the important members of transition metal dichalogenides which is emerging as a potential 2D atomically thin layered material for low power electronic and opto-electronic applications. However, for MoS$_2$ a critical fundamental question of significant importance is how the surface energy and hence the wettability is altered in nanoscale -- in particular, the role of crystal quality in low dimensions. Present work reports the synthesis of large area MoS$_2$ films on insulating substrates with different surface morphology via vapor phase deposition by varying the growth temperatures. The crystallinity of the samples is examined by transmission electron microscopy and Raman spectroscopy. From contact angle measurements, it is possible to correlate the wettability with crystallinity at nanoscale. The specific surface energy for few layers thick MoS$_2$ is estimated to be around 46.5 mJ/m$^2$. Our results shed light on the MoS$_2$-water interaction which is significant for developing important devices based on MoS$_2$ coated surfaces for micro-fluidic applications.**

--------------

*rkatiyar@hpcf.upr.edu, satya504@gmail.com

Since the discovery of graphene, two-dimensional [2D] materials have been receiving tremendous research interest. Although graphene exhibits excellent physical properties,[1,2,3,4,5,6] its application in future electronic devices is limited by its zero bandgap[7,8]. In developments parallel to that of graphene, transition metal dichalcogenides (TMDC) abbreviated as $MX_2$ where M is a metal and X is chalcogen (S, Se or Te) are emerging as alternative [2D] materials for various applications[6,9,10,11,12,13]. In each layer of TMDC the M and X atoms are bonded covalently in a prismatic arrangement, and layers are held together by van-der Waals forces[14], making it possible to obtain a single layer by mechanical or liquid exfoliation methods[15,16]. Among these TMDCs recently monolayer $MoS_2$ has been studied extensively due to its excellent electronic and optical properties[17,18,19]. For instance, in bulk it has an indirect band gap (1.2 eV) whereas it changes to direct band gap in monolayer (1.8 eV). Kim *et. al.* have demonstrated excellent room-temperature mobility in monolayer $MoS_2$ for its application in low-power nano-electronics and optoelectronics[20]. Although there is considerable progress in understanding the electronic, optical, and spintronics properties of $MoS_2$, less effort has been made to understand the wettability of $MoS_2$ by studying the water-$MoS_2$ interaction. The wettability property of atomically layered $MoS_2$ is extremely important from both fundamental and application points of view. It has been demonstrated that the presence of humidity can alter the $MoS_2$ transistor characteristics[21]. Moreover, [2D] materials are equally important for coatings, micro fluids, cell proliferation, performance of micro/nano electronics and in hybrid structures[22,23,24,25]. But to serve these purposes, intrinsic surface properties such as adhesivity, hydrophobicity, wettability, and environmental compatibility are important, and these properties depend upon the chemical composition, crystallinity and topology of the surfaces. $MoS_2$ has been used as catalyst for photo induced $H_2$ evolution and hydro-sulphurization in form of nano particles, tubes, and in forms loaded with co-catalysts[26, 27, 28]. For such applications recognition of active sites and surface hydrophilicity (i.e., wettability) is very important[29, 30, 31]. Recently contact angle measurement and molecular dynamics calculations on wettability of graphene monolayers established their hydrophobic as and translucent nature, whereas the hydrophobic nature of graphene depends on the surface energy (i.e., minimum surface energy will produce maximum hydrophobic nature and vice versa[32, 33, 34]).

Recently edge and basal-plane oriented $MoS_2$ films were synthesized via rapid sulphurization of Mo films at different temperatures which could be utilized to study the wetting property of $MoS_2$ films[35, 36]. The present work involves the growth of large-area $MoS_2$ thin films onto insulating substrates at different synthesis temperature by sulphurization of metallic Mo film. We employed high resolution transmission electron microscopy (HRTEM) to determine the microstructure of the films. Whereas optical absorption, transmission and vibrational spectra reveal the crystalline nature of the films at grown at different temperatures, x-ray photoelectron microscopy (XPS) reveals the 2H-$MoS_2$ phase. Detailed microstructure of these films is correlated with the wettability of $MoS_2$ via static contact angle (CA) measurement at room temperature. Static CA was measured for different liquids on $MoS_2$ films grown at 900 °C, and CA values utilized to determine the surface energy using Neumann's method. Film preparation and measurements are given in method section in detail.

**Result and discussion:**

Figure 1(a) shows the selected area electron diffraction (SAED) image of $MoS_2$ layer prepared at 550 °C. The diffuse ring pattern indicates the quasi-crystalline nature of the film. The HRTEM image 1(b) reveals some peculiar features: apart from the characteristic [2D] growth pattern there are several fringe-like pattern. The separation measured between two consecutive fringes is ~0.6-0.7 nm, which matches well with the known MoS$_2$ interlayer separation. Therefore these fringe-like features represent the vertically aligned edge-terminated structure of MoS$_2$ whose occurrence in a given area is almost the same as those of the flat regions. The length and width of each edge-terminated structure varies from 5-7 nm and 3-5 nm, respectively. In the enlarged HRTEM image (Fig 1(c)) the highlighted regions show amorphous (I), nano crystalline (II) and vertically aligned (III) areas and their corresponding Fast Fourier transformation (FFT) images. The FFT image of region II exhibits perfect hexagonal structure, and confirms the crystalline nature of the flat region.

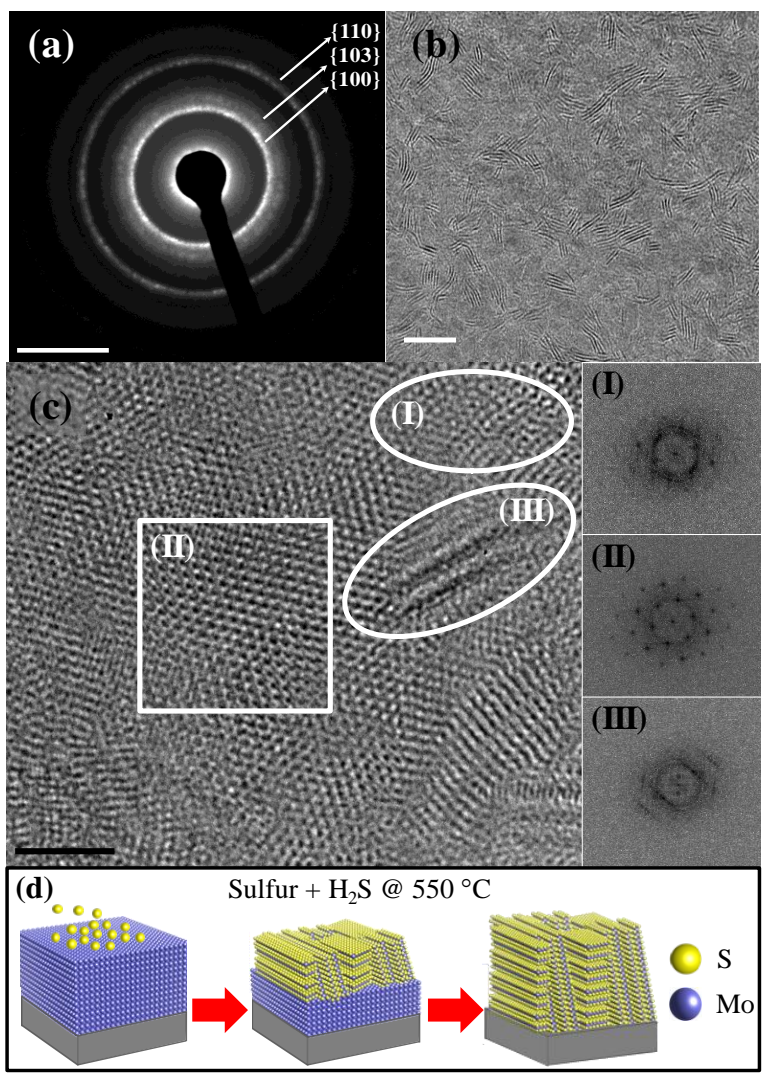

**Figure 1: MoS$_2$ growth at 550 °C – Structural characterization: (a) SAED image (b) vertically oriented MoS$_2$ layers (c) enlarged HRTEM image with (I) amorphous, (II) crystalline and (III) vertically oriented regions respectively. Inset shows their FFT and (d) Schematic of growth mechanism for vertically aligned MoS$_2$ layer. Scale bar in (a) 5 nm$^{-1}$,(b) 5 and (c) are 2 nm, respectively.**

Since the surface energy of the edge-terminated structure is considerably higher compared to that of the flat [2D] basal-plane structure of MoS$_2$ (almost two orders of magnitude higher[37]), it is often considered as a metastable structure[38]. In a recent study a similar result has been reported by Kong et al (ref 35). In general, layered materials such as graphene, hexagonal boron nitride and metal chalcogenides preferably grow in [2D] basal planes in contact with the surface of the substrate; and further layers add up to form the horizontal stacked structure which minimizes the surface energy. Although there are exceptions, such as carbon nanotubes, fullerene, etc., Kong et al. have observed the dominance of edge-terminated structures in their experimental conditions. They have explained the growth mechanism by considering the kinetic process of sulfur diffusion: during low-temperature synthesis, due to the structural anisotropy, the rate of sulfur diffusion across the layer gap is much more rapid than diffusion through the layers, resulting in vertically oriented structures. The key difference in the synthesis process in the present study as compared to Kong et al. is the sulfurization of Mo film in a reducing atmosphere of H$_2$+Ar. At elevated temperature (at about 450 °C) sulfur reacts with H$_2$ and converts into H$_2$S which reacts with Mo film faster than sulfur diffusion does. These two process of chemical conversion compete and result in both vertically aligned edge-terminated structure (due to sulfur diffusion) and regular [2D] stacked/platelet structures (chemical conversion of MoS$_2$ due to H$_2$S).

Crystallinity improves further for the sample prepared at 750 ºC which can be confirmed from the SAED image shown Fig. 2(a). Figure 2(b) shows two highlighted regions (in white). Region 1 shows the grain boundary where two grains with (102) planes meet, as shown in the enlargement in figure 2(c). Region 2 shows a Moire's pattern which forms due to stacking of hexagonal layers at an angle.

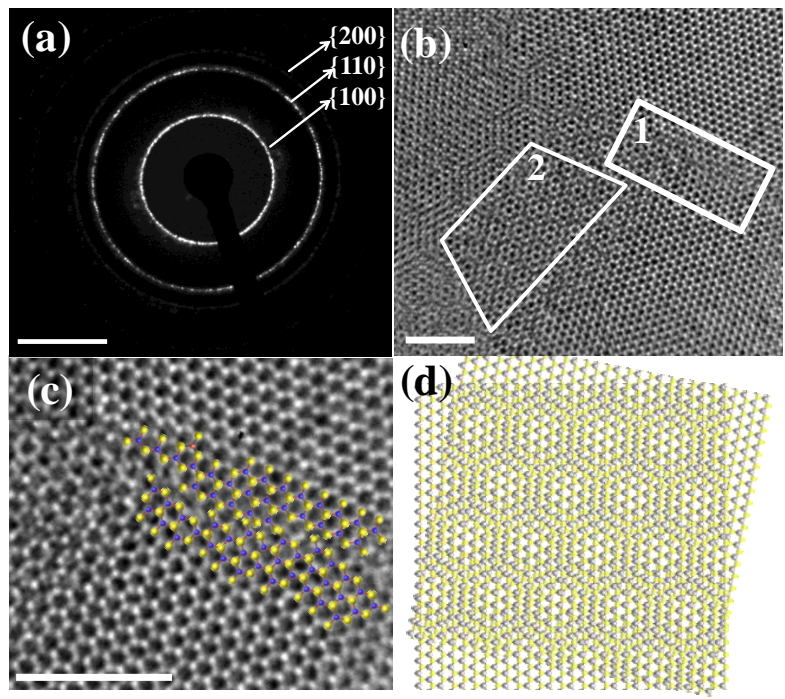

**Figure 2: MoS$_2$ growth at 750 deg C – Structural characterization: (a) SAED image of MoS$_2$ grown at 750 ºC; (b) HRTEM image of 2H-MoS$_2$ with moire pattern; (c) enlarged image of highlighted region 1 showing grain boundary of two planes; (d) simulated Moire's pattern shown for highlighted region 2. Scale bar in (a) is 5 nm$^{-1}$; 2 nm in (b) and (c).**

Using FFT analysis for highlighted region 2 reveals that layers stack upon each other at ~11°. Figure 2(d) show the simulated Moire's pattern of hexagonal $MoS_2$. However, vertically aligned layers are absent from this sample. $MoS_2$ film grown at 900 °C shows a high crystallinity with a honeycomb lattice of $2H-MoS_2$. Figure 3 (a) and (b) shows the SAED pattern and enlarged HRTEM image of hexagonal $MoS_2$ grown at 900 °C.

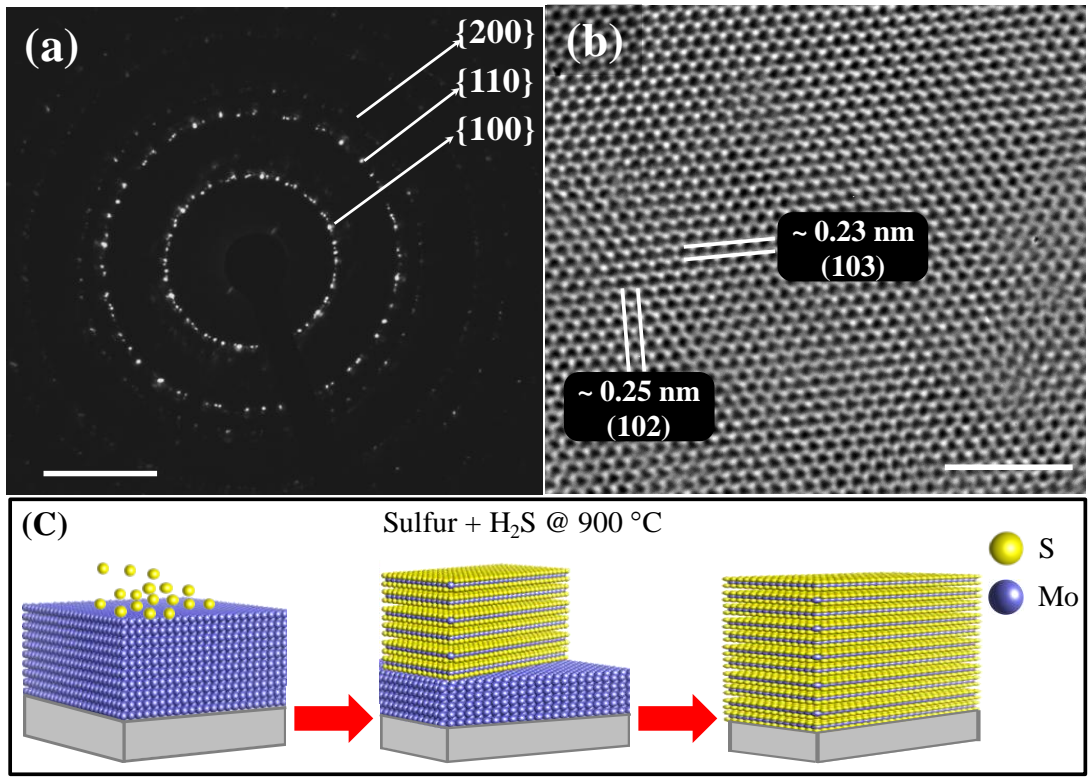

**Figure 3: $MoS_2$ growth at 900 °C – Structural characterization: (a) SAED image of $MoS_2$ grown at 900 °C; (b) hexagonal lattice of $MoS_2$ and (c) schematic presentation of basal plane growth of $MoS_2$ .Scale bar in (a) is 5 $nm^{-1}$ and in (b) is 2 nm.**

As discussed earlier, for low-temperature synthesis processes, chemical conversion due to sulfur diffusion and $H_2S$ reaction results in vertically aligned edge-terminated and regular [2D] stacked/platelet structures, respectively. However at elevated temperatures sulfur diffusion is the rate-limiting process and chemical conversion due to H₂S is much faster than the diffusion of sulfur into the film. This results in pure [2D] stacked-platelet structures without any vertically aligned edge-terminated structure.

TEM results indicate the presence of a single phase of hexagonal 2H-MoS$_2$, which is also confirmed by XPS and Raman spectra. XPS spectra for Mo(3d), S(2s), and S(2p) regions for a sample grown at 900 °C are shown in Figure 4. The Mo(3d) consists of peaks at around 229 and 232 eV belonging to Mo$^{4+}$ d$_{5/2}$ and Mo$^{4+}$ d$_{3/2}$ components respectively, and the S(2p) region has a doublet peak 2p$_{1/2}$, and 2p$_{3/2}$, which appear at 163 and 161.9 eV, respectively and confirm the -2 oxidation state of sulfur[39]. XPS spectra for other samples reveal similar information.

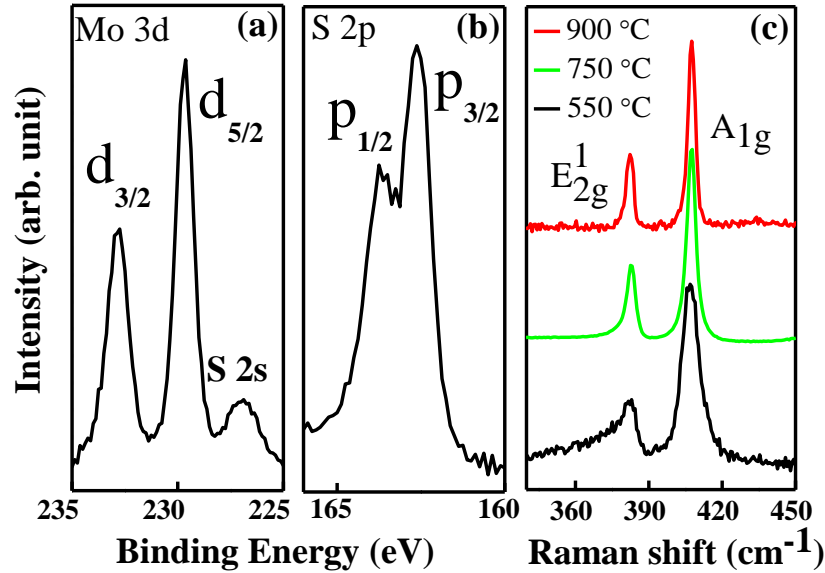

**Figure 4: Spectroscopic characterization of MoS$_2$ grown at different temperatures (a), (b) XPS for MoS$_2$ grown at 900 °C at Mo and S energies and (c) Raman spectra of MoS$_2$ prepared at 550 °C, 750 °C and 900 °C.**

Raman spectroscopy is one of the most powerful nondestructive tools to study [2D] layered materials particularly to estimate the number of layers (accurately up to seven layers), defects and electron-phonon interaction. The crystal lattice of the bulk $MoS_2$ belongs to the $D_{6h}$ point group symmetry with four Raman active modes, and these zone centre phonon modes can be represented by the following irreducible representations: $\Gamma=A_{1g}+2E_{2g}+E_{1g}$. Figure 4 (c) shows the room-temperature Raman spectra of $MoS_2$ prepared at three different reaction temperatures on $SiO_2$/Si. All samples have two distinct peaks at about 382 and 409 cm$^{-1}$ which are assigned to $E^1_{2g}$ and $A_{1g}$ modes, respectively[40, 41]. $E_{1g}$ mode is forbidden in a backscattering geometry, for the plane perpendicular to the c-axis and we do not observe this mode for any of our samples. The frequencies of vibrations of these phonon modes depend on the number of layers, and a systematic change in the lineshape parameters (both in peak position and width) have been observed[42]. As can be seen from Fig. 4(c), both the $E_{1g}$ and $A_{1g}$ modes are significantly broader for the sample prepared at 550 $^o$C, and the broadening decreases for samples prepared at high temperatures. There are factors such as crystallinity, particle size and defects that can significantly affect the Raman linewidths[43]. Poor crystallinity and defects can result broadening of Raman peak due to the loss of long range periodic ordering in the crystal lattice. On the other hand, particle size effects cause broadening of Raman peak due to the confinement of zone center optical phonons in small nanocrystals. As discussed in the TEM study for the sample prepared at 550 $^o$C, the length and width of each edge-terminated structure varies from 5-7 nm and 3-5 nm, respectively. This length scale is sufficiently small for optical phonons to be confined and thus results in broadening of Raman peaks.

We employed UV-visible absorption spectroscopy to investigate further the optical quality of samples. It's obvious from the digital photograph (see supporting information Fig. 1) of pure $Al_2O_3$, and $MoS_2$ prepared at 550 and 900 °C that the substrate deposited with $MoS_2$ layers are almost transparent to visible light. $MoS_2$ is an indirect band gap semiconductor with a band gap of 1.2 eV. Its electronic spectrum consist of several absorption thresholds which have been assigned as A, B, C, D and E whose absorption threshold appear at 672, 621, 447,408 and 45 nm, respectively. The first set of absorption thresholds (A and B) are excitonic transitions whose energy separation is 180 meV[44]. In the monolayer honeycomb lattice of $MoS_2$ each Mo is sixfold coordinated and sandwiched between two threefold coordinated S atoms in a trigonal prismatic arrangement. Spin-orbit coupling splits the top of the valence band in to two sub-bands at the K point of the Brillouin zone. The other excitonic transitions; C, D and E, are associated with transitions to the conduction band from deeper levels associated with the valence band. The absorbtion and transmittance spectra for sample prepared at 550 and 900 °C (see supporting information Fig. 2) reveals that excitonic peaks are very weak and poorly resolved in case of sample grown at 550 °C but these peaks are very well defined and clearly resolved for the sample grown at 900 °C. The relative intensity of the excitonic peaks depends on the crystal quality of $MoS_2$, and these absorption results are consistent with TEM and Raman studies.

**Wettability behavior:** The free energy of any surface determines the extent of spontaneous interaction with liquid or vapor and completely depends on its surface characteristics (i.e., surface topography, chemical composition, roughness etc[45, 46]). The observed difference in film characteristics of $MoS_2$ grown at temperatures 500 °C, 750 °C and 900 °C is manifest in the extent of their hydrophobic nature. Hence we performed static CA (contact angle) measurement for all three films with water.

Results are found analogous to their surface morphology as follows: CA for the sample grown at 550 ºC was found to be ~23.8º (Fig 5a), confirming the hydrophilic nature. This could be explained on the basis of its semicrystalline nature and the presence of (001) planes vertically oriented to the substrate. Wettability is a function of the specific free energy for any given surface (i.e., higher surface free energy causes increased wettability). The presence of defects, foreign contaminants, and varied chemical composition can contribute to enhance the free energy and result in higher wettability. Anisotropy in $MoS_2$ bonding results in active edge sites; therefore film surfaces containing vertically oriented structures have active sites exposed which in turn increase the surface free energy. Another contribution to surface free energy comes from the grain boundaries that exist in flat (100) region. As mentioned in the TEM discussion, the existence of nano-grain with amorphous regions might also have a significant role in enhancing the surface energy. Since water is a polar solvent, has a high surface energy ~72mJ/m$^2$, and readily reacts with active edge sites, it forms thiomolybdates (Fuerstenau and Chander) which is hydrophilic in nature[47]. However, the $MoS_2$ basal plane is chemically inert and shows a hydrophobic nature.[30] Earlier work[48] of wettability on cleaved $MoS_2$ has shown static CA ~80º, which is well below that found in present work (~ 90º) and could be due to defects created by cleaving the $MoS_2$ crystal. As shown in the HRTEM images, films grown at 900 ºC have perfect hexagonal crystallinity and expose (001) planes which are hydrophobic. Figure 5(b) shows the variation in CA with time where the dash-dot lines separate the advancing, pinning and receding regions. Contact angles were measured by fitting the drop shape utilizing active B spline contours (Image J Version 1.46). The base radius of each droplet remained almost unchanged for long times (~400 sec); hence the contact line was pinned, which in turn reduces the CA values marked at the pinning region. The measured value of advancing ($\theta_a$) and receding contact ($\theta_r$) angle were 91±1° and 68±1°. Therefore wettability shows a contact angle hysteresis (defined as difference between $\theta_a$ and $\theta_r$) which implies surface macroscopic roughness or heterogeneity. As shown in the HRTEM images, MoS$_2$ films show a polycrystalline nature that might contain some line defects at the surface and produce hysteresis in CA. These line defects hinder the motion of the contact line. At t=0 these domains prohibit the motion of the water edge, and the contact angle CA achieves a maximum value. When the liquid evaporates due the pinning of the boundary line we observe a decrease in the CA value. Eventually the droplet evaporates in such a way that both the contact angle and base line decrease simultaneously and achieve a constant value of CA. For better understanding of how the wettability/contact angle/water droplet shape change with time for the above three discussed samples, we have provided small movies in the supporting information. Note that for clarity, these movies are shown at slow speed (0.125X speed).

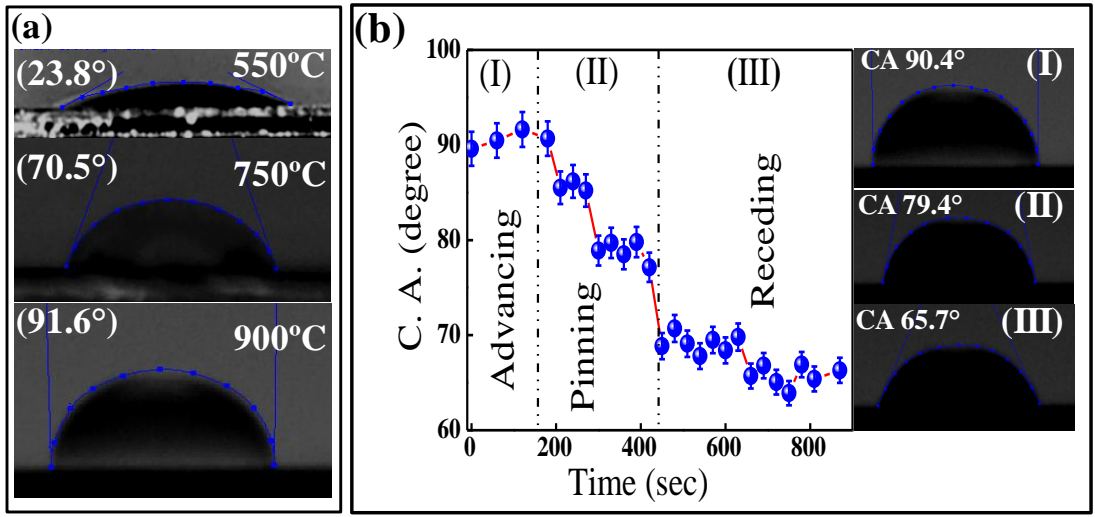

**Figure 5: Wetting angle measurement for MoS$_2$ films. (a) Static contact angle measured for MoS$_2$ grown at 550, 750 and 900 °C with water. (b) Graph shows the change in static contact angle with time for MoS$_2$ sample grown at 900 °C. Advancing, pining, and receding regions are separated by dotted lines and the photograph of water droplet at corresponding regions are also shown in the right hand side of Fig 5 (b).**

Wettability for MoS$_2$ sample grown at 900 °C is also measured for different liquids (given in table 1) with different surface tension. Due to different surface tension and polarity, the degree of wettability changes and results in different static CA values. Wetting of a solid surface by a liquid may be used to approximate the surface free energy of the solid by measuring the angle between solid and liquid. Young's equation relates these parameters under the mechanical equilibrium of liquid droplet and given as:

$$\gamma_s = \gamma_{sl} + \gamma_l \cos\theta \tag{1}$$

where $\theta$ is the contact angle between the liquid and solid surface; $\gamma_s$, $\gamma_{sl}$ and $\gamma_l$, represent the solid surface free energy, liquid surface free energy, and solid-liquid interfacial energy. Exact measurement of the solid-surface free energy is impossible by measuring the contact angle and utilizing the equation above, although several approximation methods[49, 50] exist to estimate $\gamma_s$. Neuman's equation of state theory is more reasonable for determining the surface energy and states that the interfacial free energy $\gamma_{sl}$ is a function of $\gamma_s$ and $\gamma_l$:[51, 52]

$$\cos\theta = -1 + 2\sqrt{\frac{\gamma_s}{\gamma_l}}\, e^{-\beta(\gamma_s - \gamma_l)^2} \tag{2}$$

where β is a constant with dimensions $(mJ/m^2)^2$. Although this equation is not useful for high energy solid surfaces, the energy for hydrophobic surface can be estimated to a good approximation by measuring the static contact angle for different liquids. Equation 2 can be developed further and rewritten as

$$\ln\left[\gamma_l\left(\frac{1+\cos\theta}{2}\right)^2\right] = -2\beta(\gamma_S - \gamma_L)^2 + \ln\gamma_S \quad (3)$$

Table 1 provides the information about the liquids used and their surface tension, measured contact angle values and calculated values [left hand side (LHS) of equation 3].

**Table 1: Measured Contact Angle of different liquids on the MoS$_2$ surface**

| Liquid | Ethyl alcohol | Di-ethylene glycol | Ethylene glycol | Glycerol | Water |
|---|---|---|---|---|---|
| **Surface tension @ 25 ºC** | 22.7 | 44.4 | 47.3 | 63.75 | 72.1 |
| **CA(º)** | 23.4 | 53 | 60 | 93 | 91.6 |
| $\ln\left[\gamma_l\left(\frac{1+\cos\theta}{2}\right)^2\right]$ | 3.038 | 3.34 | 3.28 | 2.66 | 2.9 |

Figure **6** shows a graph of the left-hand side of Eq.3 versus $\gamma_l$, the obtained result is fitted to the following parabolic equation,

$$y = -9.08*E^{-4} X^2 + 0.0812X + 1.509 \quad (4)$$

which gives the surface free energy of MoS$_2$ at room temperature as 44.5 mJ/m$^2$, very close to surface energy of graphene films[53].

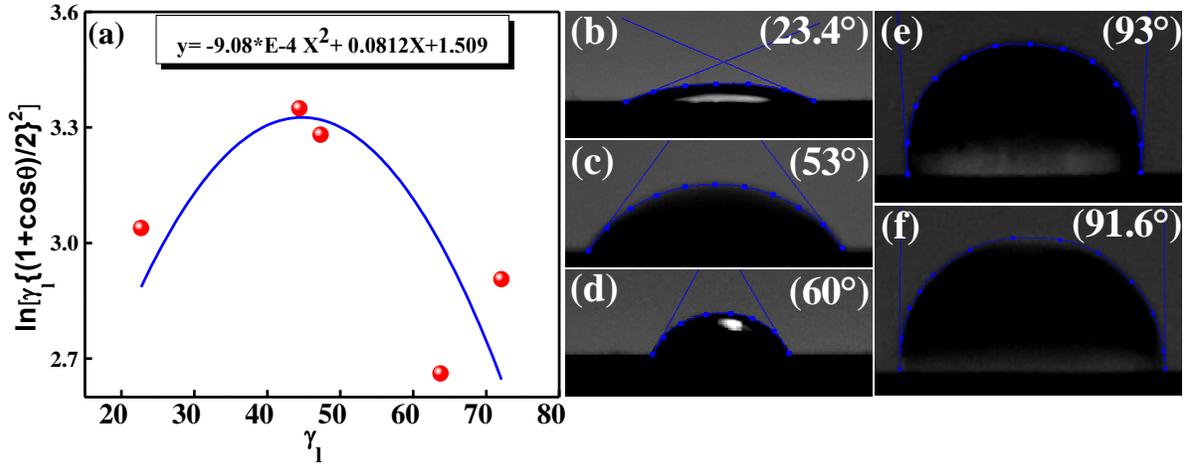

**Figure 6:** Contact angles with different liquids on $MoS_2$ grown at 900 ºC. (a) Plots of $\ln\left[\gamma_l \left(\frac{1+\cos\theta}{2}\right)^2\right]$ vs $\gamma_l$ for different liquid. Image of droplet with contact angle value for (b) Ethanol, (c) Di-ethylene glycol, (d) ethylene glycol, (e) glycerol, (f) water.

**Conclusion:** In conclusion we have synthesized large area $MoS_2$ films a few atomic layers thick on insulating substrates by vapor deposition methods at different temperatures. From the static contact angle measurements it is found that the wettability of $MoS_2$ evolves from hydrophilic to hydrophobic as the growth temperature is increased. The difference in the wettability and hence the surface energy of these films is attributed to defects, as confirmed from TEM and Raman studies. Finally, the surface energy of a highly crystalline $MoS_2$ film of several atomic layers is estimated to be 46.5 mJ/m$^2$, which is comparable that of graphene.

**Methods:**

Few layer $MoS_2$ (FLMS) were prepared in two steps. First we coated $SiO_2$/Si and $Al_2O_3$ substrates with molybdenum using magnetron sputtering technique followed by sulfurization of metallic films. Sulfurization was carried out in a tube furnace using sulfur powder as a source in a reducing atmosphere containing argon and hydrogen gases in the ratio of 9:1 at three different reaction temperatures: 550, 750 and 900 °C. The furnace temperature was raised at 10 °C/minute, and reaction time was kept at 20 minutes for all samples. Etching the $SiO_2$ using a KOH (30%) solution produced suspended FLMS that were placed on TEM copper grids without carbon film. The samples were characterized by using a TEM (Carl Zeiss LEO 922) and a high resolution TEM (HRTEM, JEOL JEM-2200FS). Raman measurements were done using a Horiba-Jobin T64000 (triple mode subtractive) micro-Raman system in a backscattering configuration. Polarized radiation from a diode laser (532 nm excitation wavelength) was focused on the sample using an 80X objective, and the laser spot size on the sample was about 1 micrometer. Laser power was minimized (0.5 mW) to avoid possible laser heating of the sample. Wetting behavior of these films was determined by measuring the static contact angle (CA) using deionized water at $MoS_2$ surface, and all the samples were heat treated at 200 °C for 30 minutes in Argon before the measurement.

**Acknowledgement: Acknowledgement:** The authors acknowledge financial support from DOE (grant DE-FG02-ER46526). A. P. S. G. acknowledges an NSF fellowship (grant NSF-RII-1002410). We thank Dr. Esteban Fachini for the XPS measurements and Materials Characterization Centre (MCC-UPR) for providing XPS facility. We also thank Prof. J. F. Scott (University of Cambridge, UK) for useful discussion in the manuscript.

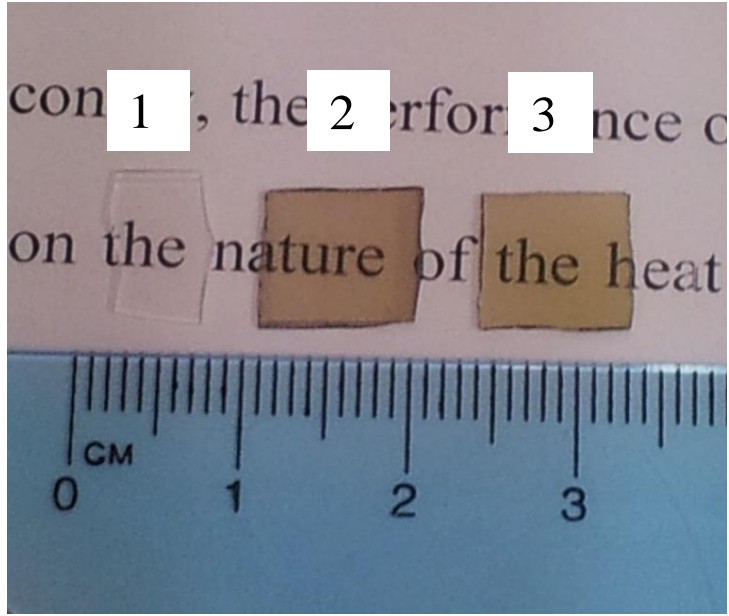

**Figure 1: Digital photograph of (1) Al$_2$O$_3$ both side polished substrate, MoS$_2$ prepared at (2) 900 ºC (3) 550 ºC**

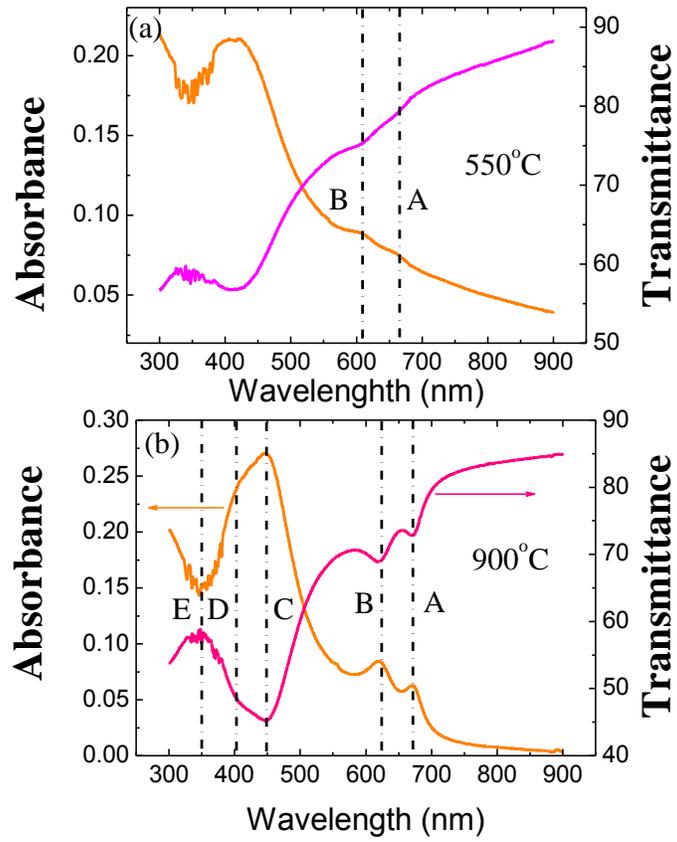

**Fig 2 UV-VIS absorption and transmission spectrum of MoS$_2$ on 2SP Al$_2$O$_3$ (a) 550 ºC (b) 900 ºC**